\documentclass[preprint,aps,prb,showpacs,groupedaddress]{revtex4}

\usepackage{amsmath}
\usepackage{dcolumn}
\usepackage{epsfig}
\usepackage{graphicx}
\usepackage{latexsym}

\usepackage{epstopdf}

\begin{document}
\hyphenation{Ka-pi-tul-nik}

\title{Feedback Control and Characterization of a Microcantilever Using Optical Radiation~Pressure}

\author{David M. Weld}
\email{dweld@stanford.edu}
\affiliation{Department of Physics, Stanford University, Stanford, CA 94305}
\author{Aharon Kapitulnik}
\affiliation{Department of Applied Physics, Stanford University, Stanford, CA 94305} 
\affiliation{Department of Physics, Stanford University, Stanford, CA 94305}

\begin{abstract}
We describe a simple method for feedback-regulation of the response of a microcantilever using the radiation pressure of a laser.  A modified fiber-optic interferometer uses one laser to read out the position of the cantilever and another laser of a different wavelength to apply a force that is a phase-shifted function of that position.  The method does not require a high-finesse cavity, and the feedback force is due solely to the momentum of the photons in the second laser.  The feedback phase can be adjusted to increase or decrease the microcantilever's effective quality factor $Q_{\mathrm{eff}}$ and effective temperature $T_{\mathrm{eff}}$.  We demonstrate a reduction of both $Q_{\mathrm{eff}}$ and $T_{\mathrm{eff}}$ of a silicon nitride microcantilever by more than a factor of 15 using a root-mean-square optical power variation of $\sim$2~$\mu$W.  Additionally, we suggest a method for determination of the spring constant of a cantilever using the known force exerted on it by radiation pressure.

\end{abstract}


\maketitle

It is useful for many applications to be able to control the effective quality factor $Q_{\mathrm{eff}}$ and temperature $T_{\mathrm{eff}}$ of a micromachined cantilever.  A system that phase-shifts the cantilever's own thermal fluctuations and feeds them back to the cantilever as force has been shown to be capable of both increasing \cite{tamayo} and decreasing \cite{mertz} the effective quality factor and temperature, as well as modifying the effective spring constant.\cite{aoki}  For scanning probe force microscopes, $Q$-modification of both signs can be useful.  For general cantilever-based force-detection experiments, reduction of $Q_{\mathrm{eff}}$ and $T_{\mathrm{eff}}$ can improve experimental convenience without sacrificing high force sensitivity.\cite{frogland2} The force that is fed back to the cantilever can be of any type; previously, this force has most often been applied using piezoelectric elements,\cite{frogland2} magnetic coatings,\cite{tamayo} or the photothermal forces that result from bimorph-type cantilevers being locally heated by a laser.\cite{mertz}   

In this letter we present a method of using radiation pressure to apply the feedback force.  This scheme is simple and robust; it has similar effectiveness to existing methods, but has the advantage of not requiring that the cantilever be modified by addition of a metallic coating or piezoelectric stack.  It is especially easy to add this capability to systems that read out cantilever position optically, since it can make use of the same focusing and alignment optics. Radiation pressure has previously been used to actuate micromechanical oscillators in several experiments.  Marti \emph{et al.} performed an early experiment investigating the mechanical effects of radiation pressure on micromachined cantilevers.\cite{marti}  A high-sensitivity atomic force microscope has been developed for biological applications that uses optical radiation pressure to control the position (but not the quality factor) of a force-sensing cantilever in liquid.\cite{aoki}  Acoustic radiation pressure has been shown to be an effective tool for actuation and characterization of microcantilevers in fluids.\cite{degertekin}  At larger length scales, an elegant proof-of-principle experiment designed to test the technology for gravity-wave interferometers showed that radiation pressure can be used to control the oscillations of a macroscopic mirror.\cite{cohadon}  The present work extends the results of those experiments by demonstrating a simple method of controlling the quality factor and temperature of a microcantilever using only optical radiation pressure.

In general, the displacement $x$ of a damped harmonic oscillator as a function of frequency $\omega$  is 
\begin{equation}
x(\omega) = \frac{\omega_{\mathrm{o}}^2/k}{\omega_{\mathrm{o}}^2 - \omega^2 + i\Gamma\omega} [F_{\mathrm{thermal}}(\omega) + F_{\mathrm{ext}}(\omega)],
\end{equation}
where $k$ is the spring constant, $\omega_{\mathrm{o}}$ is the resonant frequency, and $\Gamma=\omega_{\mathrm{o}}/Q_{\mathrm{o}}$ is the intrinsic damping of the oscillator.  Here, $F_{\mathrm{thermal}}(\omega)$ represents the random thermal Langevin force and $F_{\mathrm{ext}}(\omega)$ an externally applied force, which in this case is due to radiation pressure.  The applied force can be modulated by a feedback loop whose input is the measured displacement.  Adjusting the phase of the feedback gain at the resonant frequency to $\pi/2$ has the effect of producing a velocity-dependent force at the resonant frequency.  In particular, if the gain is chosen so that the applied force near resonance is
$F_{\mathrm{ext}} = -i m \omega g  x$,
where $m$ is the mass of the oscillator and $g$ is proportional to the magnitude of the feedback gain on resonance, then the displacement as a function of frequency becomes
\begin{equation}
x'(\omega) = \frac{\omega_{\mathrm{o}}^2/k}{\omega_{\mathrm{o}}^2 - \omega^2 + i[\Gamma+ g]\omega} [F_{\mathrm{thermal}}(\omega)].
\end{equation}
Assuming that the noise of the feedback system can be neglected, the feedback thus changes the damping of the system without adding fluctuations.  This leads to a changed effective quality factor $Q_{\mathrm{eff}}=\omega_{\mathrm{o}}/[\Gamma + g]$ and a changed effective temperature $T_{\mathrm{eff}}=T_{\mathrm{o}} \Gamma/[\Gamma + g]$, where $T_{\mathrm{o}}$ is the temperature of the oscillator's environment.\cite{cohadon}  A positive $g$ lowers both $Q$ and $T$ by the same factor.

\begin{figure}[ht]
\begin{center}
\includegraphics[width=0.85 \columnwidth]{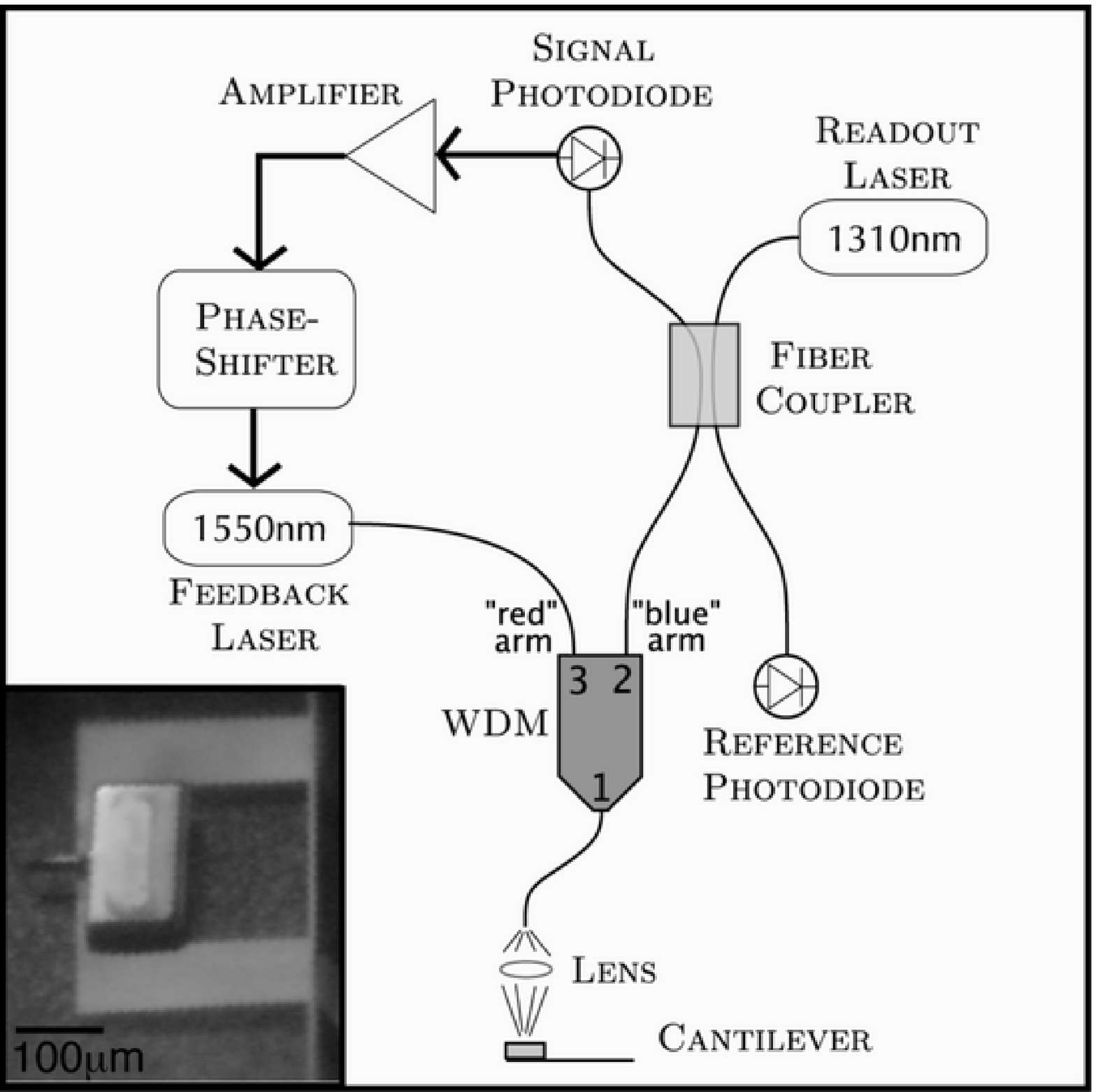}
\end{center}
\vspace{-4mm}
\caption{ Experimental setup; see text for details.  Inset photograph shows the cantilever with a gold mass attached.} 
\label{setup}
\end{figure}

The apparatus used for demonstrating feedback cooling of a cantilever with radiation pressure is depicted in figure~\ref{setup}.  It consists of a modified fiber-optic interferometer of a type first proposed by Rugar \emph{et al}.\cite{rugar-intf}  A 0.5~mW 1310~nm distributed-feedback diode laser (PD-LD PL13U0.51FAB-T-1-01)injects light into a single-mode optical fiber (Thorlabs 1060XP).  The light travels through a standard 99/1 fiber coupler and into the ``blue'' arm of a cascaded wave division multiplexer (2$\times$ JDSU FFW-4C6P1103), then exits the fiber through a flat cleave and is focused by an aspheric lens (Lightpath 350450C) onto a gold mass on a microcantilever.  The reflected light from the cantilever interferes with the reflected light from the cleaved end of the fiber. This results in a total reflected power that depends on the cantilever's position as $P_{\mathrm{out}}=P_{\mathrm{o}} (1-V\cos 4\pi d / \lambda)$, where $\lambda$ is the wavelength of the laser, $d$ is the distance from the cleaved fiber end to the cantilever, $P_{\mathrm{o}}$ is the midpoint power, and $V$ is the fringe visibility.\cite{rugar-intf}  A photodiode (UDT FCI-INGAAS-100L-FC) attached to another arm of the fiber coupler produces a current proportional to the reflected power; this current is then converted to a voltage by a transimpedance amplifier with a 10~M$\Omega$ feedback resistor.  This voltage is phase-shifted by a custom-built analog circuit and used to modulate the power of a 1.5~mW 1550~nm diode laser (Thorlabs S1FC1550).  Because photon momentum can only apply force in one direction, it is necessary to add a constant offset to the power so that the force modulation can be of either sign.  The light from the 1550~nm laser is added to the fiber by coupling through the ``red'' arm of the wave division multiplexer (WDM); it then follows the same optical path as the 1310~nm laser, and is focused onto the cantilever by the same optics.  The width of the focal spot is about 10~$\mu$m: much smaller than the width of the cantilever crossbar.  This makes it easy to align the lens so that all the light from the laser hits the cantilever.  In practice, alignment is achieved by temporarily replacing the 1550~nm laser with a visible laser and observing the focused spot on the cantilever through a microscope.  The WDM prevents backscattered 1550~nm light from getting to the readout photodiode; its attenuation factor is measured to be greater than 50~000.  

The custom-built cantilever used in this work was designed to be used as a force sensor in a next-generation test of Newtonian gravity at length scales of 20~$\mu$m.\cite{frogland2}  It is 230~$\mu$m long and 0.34~$\mu$m thick, and its U-shaped construction gives it an effective width of 180~$\mu$m.   In order to sense mass-dependent forces, it has a gold mass weighing approximately 5~$\mu$g attached to the end; this results in a resonant frequency of $f_{\mathrm{o}}= \omega_{\mathrm{o}}/2 \pi \simeq $ 350~Hz (implying a spring constant of $\sim$0.02~N/m). This large mass is not essential for the work described in this paper.  It is, however, convenient, not only because it provides a good reflecting surface, but also because it pushes the thermal time constant up to several seconds.  This virtually eliminates photothermal effects and Knudsen forces\cite{passian} at the resonant frequency.  The intrinsic quality factor $Q_{\mathrm{o}}$ of the cantilever is $\sim$12~000 at  $10^{-6}$~torr and  300$^\circ$K, and can be as high as 80 000 at 5$^\circ$K.

\begin{figure}[ht]
\begin{center}
\includegraphics[width=1 \columnwidth]{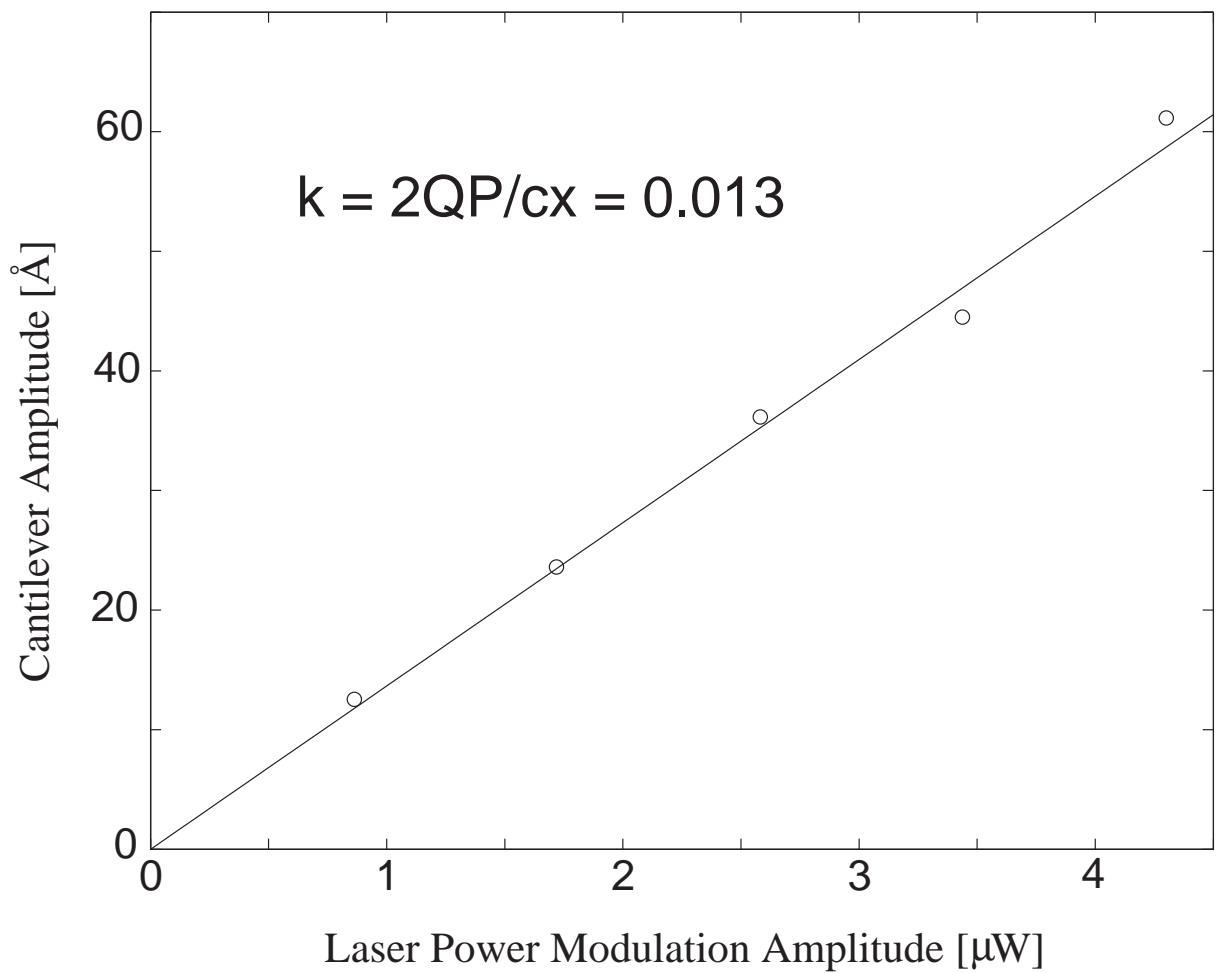}
\end{center}
\vspace{-4mm}
\caption{Cantilever displacement versus laser power (AC measurement).   The solid line is a fit to the data that indicates a spring constant of 0.013. }
\label{ampvsradforce}
\end{figure}

The radiation force exerted on a perfectly reflecting surface by a light beam of power $\mathbf{P}$ is $F_{\mathrm{rad}} = 2\mathbf{P}/c$, where $c$ is the speed of light.  To measure the force applied by radiation pressure, the power of the laser was sinusoidally modulated and the cantilever displacement was recorded as a function of the amplitude of this modulation.  Figure~\ref{ampvsradforce} shows the results of such an experiment.  The laser power was modulated at the cantilever's resonant frequency $f_{\mathrm{o}} \simeq 350$~Hz so that the motion would be amplified by a factor of $Q_{\mathrm{eff}}$, which was maintained at a value of 2700 using feedback.  Photothermal forces on the cantilever are not only too slow to have measurable effects at this frequency, but also happen to be of opposite sign from radiation pressure.  The sign and magnitude of the results are consistent with what would be expected from actuation due only to radiation pressure.  The inferred spring constant is slightly smaller than expected; this is due to the fact that the laser was focused at a point farther out along the cantilever than the center of the gold mass.  The measured value of $k$ will have a strong dependence on the exact position of the laser spot on the cantilever;\cite{sader} this position would need to be well characterized for a $k$-measurement using this technique to be accurate.  Still, since the applied force depends only upon easily measurable quantities (cantilever reflectivity and optical power), this method could furnish a new and useful way of measuring $k$.  The same experiment can also be performed at a frequency below  $\omega_{\mathrm{o}}$, in which case the (known) applied force divided by the (measured) amplitude of motion directly gives the spring constant $k$ of the cantilever without requiring knowledge of $Q$.  

The factor by which $T$ and $Q$ are reduced by feedback is proportional to the gain factor $g$ defined earlier.  The maximum value of $g$ that can be attained using a laser with a maximum rms power modulation amplitude $\langle\mathbf{P}_{\mathrm{mod}}\rangle$ is 
\begin{equation}
g = \frac{2 \langle\mathbf{P}_{\mathrm{mod}}\rangle \omega_{\mathrm{o}}}{c k \langle x \rangle} = \frac{2 \langle\mathbf{P}_{\mathrm{mod}}\rangle \omega_{\mathrm{o}}}{c \sqrt{k k_B T_{\mathrm{o}}}},
\end{equation}
where we have used the equipartition theorem to write the root-mean-square cantilever position $\langle x \rangle$ in terms of temperature $T_{\mathrm{o}}$.  It should be noted that at low temperatures, because the position fluctuations are smaller, less laser power is needed to achieve a given $g$.  For the damping experiment described here, which was done at room temperature using a maximum $\langle\mathbf{P}_{\mathrm{mod}}\rangle$ of 2 $\mu$W, the maximum value of $g$ was 4.0 $s^{-1}$, corresponding to a possible reduction in $Q$ and $T$ by a factor of $\sim$20.  

\begin{figure}[ht]
\begin{center}
\includegraphics[width=1.0 \columnwidth]{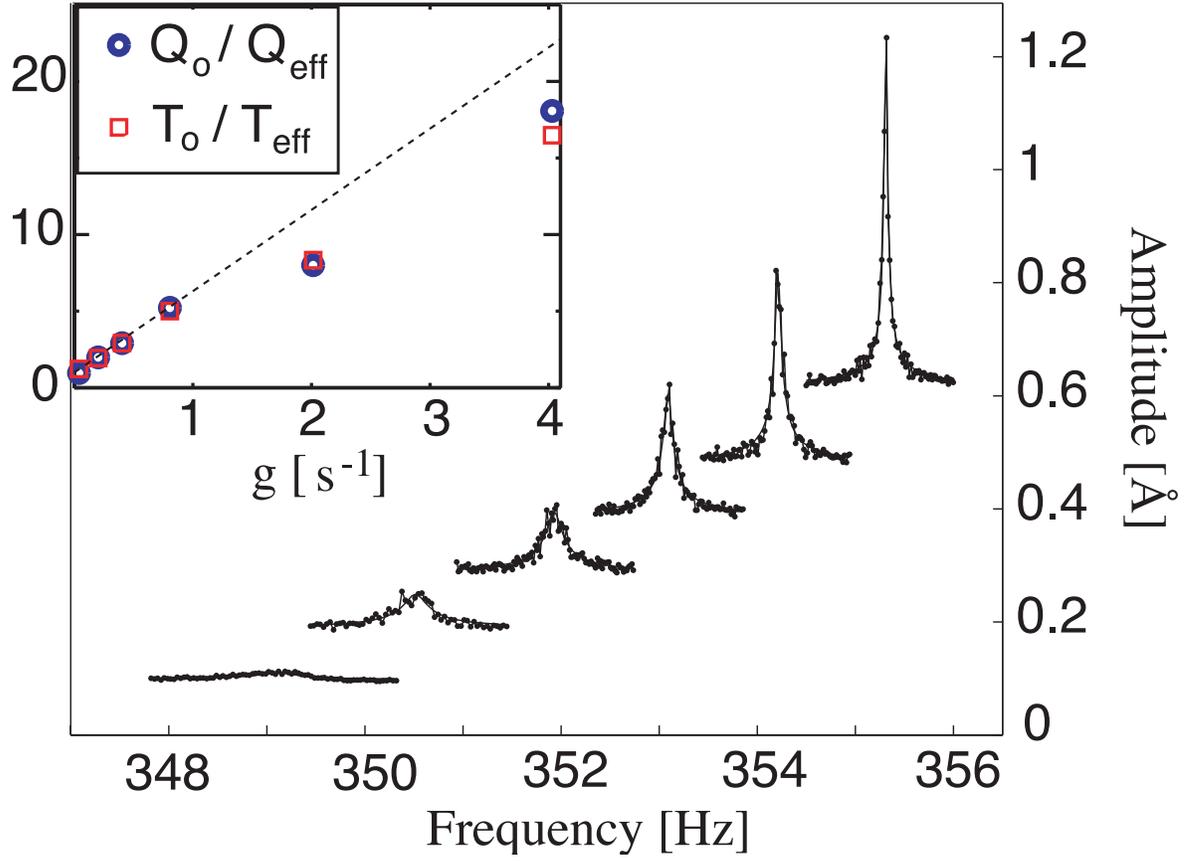}
\end{center}
\vspace{-4mm}
\caption{(Color Online) Displacement spectra taken at different gains.  Gain increases from upper right to lower left.  Peaks have been offset in $x$ and $y$ for clarity.  Inset: $Q_{\mathrm{o}}/Q_{\mathrm{eff}}$ (circles) and $T_{\mathrm{o}}/T_{\mathrm{eff}}$ (squares) versus gain factor $g$, for the same data. The dashed line is the theoretical prediction.} 
\label{spectra}
\end{figure}

Results of the feedback-modification of $Q$ and $T$ are presented in figure~\ref{spectra}, which shows the broadening and flattening of the thermally excited resonance peak with increasing feedback gain.  The individual spectra were each fitted with a Lorentzian function to extract the value of $Q_{\mathrm{eff}}$.  The effective temperature $T_{\mathrm{eff}}$ was determined by integration of the power spectral density.  Analysis of the lower-leftmost curve shows that the effective temperature of the cantilever was reduced to 18$^\circ$K, and its quality factor was reduced to $\sim$700.  The measured variation of $Q_{\mathrm{eff}}$ and $T_{\mathrm{eff}}$ with gain is shown in the inset of figure~\ref{spectra}, along with the theoretical prediction.  The agreement with theory at lower gains is excellent.  At higher gains, the performance of the feedback system becomes less ideal; this effect seems to be due to the increased importance of amplifier noise, in both the position detection and feedback amplifiers.  The maximum reduction factor achievable using this technique will likely be limited either by noise in the feedback amplifier or by cantilever heating due to the damping laser. 
 
In conclusion, we have presented a simple and robust method for controlling the effective quality factor and temperature of a cantilever using the radiation pressure of a laser.  Using this method, we have demonstrated a reduction in both  $T_{\mathrm{eff}}$ and  $Q_{\mathrm{eff}}$ by a factor of more than 15.  Additionally, we have suggested a new way to measure the spring constant of a cantilever by using a known force applied at a known location by radiation pressure.

The authors would like to thank Blas Cabrera and Dan Rugar for many helpful discussions, and Jing Xia for his work on the cantilever microfabrication.  This work was supported by NSF grant number PHY-0554170.

\pagebreak[4]


\pagebreak[4]

\end{document}